\documentclass[aps,prb,twocolumn,showpacs]{revtex4}
\usepackage{epsfig}
\usepackage{times}
\usepackage{amsmath}
\begin{document}

\title{Effects of correlated Gaussian noise on the mean firing rate and correlations of an electrically coupled neuronal network}

\author{Xiaojuan Sun}
\email[Electronic address (corr.): ]{sunxiaojuan.bj@gmail.com}
\affiliation{Zhou Pei-Yuan Center for Applied Mathematics,
Tsinghua University, Beijing, 100084, People's Republic of China
\& Division of General Mechanics, Beihang University, 100191
Beijing, People's Republic of China}

\author{Matja{\v z} Perc}
\email{matjaz.perc@uni-mb.si} \affiliation{Department of Physics,
Faculty of Natural Sciences and Mathematics, University of
Maribor, Koro{\v s}ka cesta 160, SI-2000 Maribor, Slovenia}

\author{Qishao Lu}
\email{qishaolu@hotmail.com} \affiliation{Division of General
Mechanics, Beihang University, 100191 Beijing, People's Republic
of China}

\author{J\"urgen Kurths}
\email{kurths@pik-potsdam.de} \affiliation{Potsdam Institute for
Climate Impact Research, 14412 Potsdam, Germany \& Institute of
Physics, Humboldt University Berlin, 12489 Berlin, Germany}

\begin{abstract}
In this paper, we examine the effects of correlated Gaussian noise on a two-dimensional neuronal network that is locally modeled by the Rulkov map. More precisely, we study the effects of the noise correlation on the variations of the mean firing rate and
the correlations among neurons versus the noise intensity. Via numerical simulations, we show that the mean firing rate can always be optimized at an intermediate noise intensity, irrespective of the noise correlation. On the other hand, variations of the
population coherence with respect to the noise intensity are strongly influenced by the ratio between local and global Gaussian noisy inputs. Biological implications of our findings are also discussed.\end{abstract}

\pacs{05.45.-a, 05.40.-Ca, 05.45.Ra, 87.19.lj}

\maketitle

\textbf{It is thoroughly documented and established that noise can
play a constructive role in neuronal systems. Noise correlations,
which have been observed in the brain, are thereby usually assumed
to be ignorable. However, it has been shown that noise correlations
cannot be avoided and may indeed play a vital role in neuronal
dynamics, mainly because they affect the amount of information
transmitted across the cortex as well as the computational
strategies of neuronal networks. In the context of transmission of
neuronal information, there is an ongoing debate about whether a
cortical neuron is sensitive to the mean firing rate of presynaptic
neurons and their correlations or not. Regardless of this, the mean
firing rate and the correlations among neuronal groups are two
important factors determining the transmission of information across the network. Here
we elaborate on the effects of correlated Gaussian noise (noise
correlation) on the mean firing rate and the correlations among neurons of an
electrically coupled neuronal network. We find that the noise
correlation has little effect on the variations of the mean firing
rate with respect to the noise intensity. Variations of the
population coherence with respect to the noise intensity, however,
are strongly influenced by the ratio between local and global Gaussian
noise in the overall noise intensity, which can be tuned by the
noise correlation. Our results indicate that noise correlation may
have a significant impact on the response of postsynaptic neurons if
these are sensitive to correlated neuronal activities.}

\section{Introduction}
\label{intro} Neurons are usually subject to random fluctuations on
different scales, ranging from channel noise created by random ion
flow across the plasma membrane to synaptic noise created by the
activity of other neurons. In experiments, it has been shown that
noise has constructive effects on neuronal dynamics. For example,
William and Durand~\cite{William} showed that an appropriate noise
intensity can improve the detection of subthreshold signals in a
resonant manner. Higgs et al.~\cite{Matthew} found that synaptic
noise increases the gain in many pyramidal neurons with large
sAHP (slow after hyperpolarization). Jacobson et al.~\cite{Jacobson}
found that channel noise contributes significantly to membrane
voltage fluctuations at the subthreshold voltage range. In
theoretical and computational studies, the constructive role of
noise in neuronal systems has been reported as well. For example, it was shown that noise is able to evoke coherence and stochastic resonance in single
neurons,~\cite{Pikovsky, Longtin, Collins, Kreuz_1, Kreuz_2} as well as in one-dimensional~\cite{Kwon, Wang_YQ} and two-dimensional~\cite{Busch, Perc_PRE_2005, SUN_PhyA, QY_Wang_2006} neuronal networks. Related to the present study, in the sense that correlated noisy inputs have been considered, are the two papers by Kreuz et al.~\cite{Kreuz_1, Kreuz_2} where it has been shown that an intermediate noise intensity can evoke the most coherent temporal output of a single neuron irrespective of the noise correlation length. Double coherence resonance in terms of an optimal combination of noise intensity and correlation was reported as well. Notably, the effect of auto- and cross-correlations of input spikes on the response of spiking neurons has also been studied extensively.~\cite{trotl1} Additionally, noise can also induce and/or enhance the synchronization in neuronal systems.~\cite{Wang_YQ, ShiX_2005} Refs. \onlinecite{Gammaitoni} and \onlinecite{Lindner} are two comprehensive review papers,
recommended to the readers who are interested in the research of noise effects on nonlinear systems in general, including neuronal dynamics.

In the cortex, a single interneuron can be connected with tens of
thousands of local circuit interneurons. Thus, activity of this
single interneuron will provide correlated input to many neurons
in the local circuit.~\cite{Salinas} Furthermore, correlated
inputs might be stimulus driven and usually are random. These random
correlated inputs can be described by means of correlated noise. In the past,
noise correlations have, apart from a few exceptions,~\cite{Kreuz_1, Kreuz_2} usually been ignored. But some researchers have found that noise correlations can have many different effects on the neuronal population $-$ the amount of information encoded,
the computational strategies of networks of neurons, etc., as
reviewed in Ref.~\onlinecite{Averbeck}. This means that noise
correlations should not be neglected in neuronal systems. Adding to this conclusion is also the fact that the conceptually related coupling via noise in one-dimensional~\cite{pika1} and two-dimensional~\cite{hepan1} systems has been found to induce synchronization. Meanwhile, the question of how a postsynaptic neuron is affected by the presynaptic neuronal population is still not fully understood.
There is an ongoing debate on whether a cortical neuron is driven
mainly by the mean firing rate of presynaptic neurons or by
correlated firing activities.~\cite{Abeles, Shadlen, Softky} No
matter what the outcome of the debate, the mean firing rate and
correlations of neuronal groups are two important factors in
investigating transmission of neuronal information. Therefore, we
will investigate the effects of correlated Gaussian noise (noise
correlation) on the mean firing rate and correlations of a
neuronal population in this paper. The obtained results may have important implications for understanding the transmission of neuronal
information.

The paper is organized as follows. Equations governing the
two-dimensional neuronal network are presented in the next section.
Measures used for quantifying the observed neuronal dynamics are introduced in section~\ref{sec:3}, while the results due to correlated Gaussian noise are presented in
section~\ref{sec:4}. Finally, the summary is given in section~\ref{sec:5}. We also provide an algorithmic description of noise generation in the Appendix.

\section{Equations of the network}
\label{sec:2}

The Rulkov map \cite{Rulkov_2001, Rulkov_2002} is employed to model
the dynamical behavior of neurons constituting the examined
neuronal network. The model captures succinctly main dynamical
mechanisms in real neuronal ensembles, foremost showing typical
restructuring of collective behavior following stochastic inputs.
Specifically, we consider a network of $N \times N$ electrically
coupled Rulkov maps
\begin{equation}
\begin{array}{ll}
u_{n+1}(i,j)=\alpha/[1+u^2_n(i,j)]+v_n(i,j)+D[u_n(i+1,j)\nonumber\\
~~~~~~~~~~~~~+u_n(i-1,j)+u_n(i,j-1)+u_n(i,j+1)\nonumber\\
~~~~~~~~~~~~~-4u_n(i,j)]+\eta_n(i,j),\nonumber\\
v_{n+1}(i,j)=v_n(i,j)-\beta u_n(i,j)-\gamma,\label{eq.1}
\end{array}
\end{equation}
where $u_n(i,j)$ is the membrane potential of neuron $(i,j)$ and
$v_n(i,j)$ is the corresponding ion concentration at the discrete
time $n$. The system parameters are $\alpha$, $\beta$ and
$\gamma$, whereby the latter two determine the time scale
associated with the dynamics of the slow variable $v_n(i,j)$ and
$\alpha$ is the main bifurcation parameter. If not stated
otherwise, we use $\alpha=1.99$ and $\beta=\gamma=0.001$, for
which each neuron is governed by a single excitable steady state
$(u^*,v^*)=(-1,-1-\alpha/2)$. Each neuron is coupled electrically
with its four nearest neighbors with periodic boundary conditions
given by
$u(0,j)=u(N,j),u(N+1,j)=u(1,j),u(i,0)=u(i,N),u(i,N+1)=u(i,1)$.
Finally, $D$ is the coupling strength between the neurons on the
$128 \times 128$ spatial grid.

The correlated Gaussian noise $\eta_n(i,j)$ is expressed as
\begin{eqnarray}
\eta_n(i,j)=\sqrt{R}e_n+\sqrt{1-R}\xi_n(i,j),
\end{eqnarray}
where $e_n$ is the Gaussian white noise and common to all units,
\textit{i.e.} global noise, with the properties:
\begin{equation} \left\{ \begin{array}{ll}
\langle e_n\rangle=0, \\
\langle e_ne_m)\rangle=2\sigma\delta_{m,n},
\end{array}
\right.
\end{equation}
and $\xi_n(i,j)$ is the local Gaussian noise, which is
uncorrelated from site to site. $\xi_n(i,j)$ is taken as Gaussian
white noise with the properties:
\begin{equation} \left\{ \begin{array}{ll}
\langle \xi_n(i,j)\rangle=0, \\
\langle \xi_n(i,j)\xi_m(i^{'},j^{'})\rangle=2 \sigma_{loc}
\delta_{i,i^{'}}\delta_{j,j^{'}}\delta_{m,n},
\end{array}
\right.
\end{equation}
and Gaussian colored noise with the properties:
\begin{equation}
\left\{
\begin{array}{ll}
\langle \xi_n(i,j)\rangle=0,\\
\langle \xi_n(i,j)\xi_m(i^{'},j^{'})\rangle=\sigma_{loc} \lambda
exp(-\lambda |n-m|)\delta_{i,i^{'}}\delta_{j,j^{'}},
\end{array}
\right.
\end{equation}
respectively. Here $\sigma$ is the noise intensity of the global
noise $e_n$, $\sigma_{loc}$ is the noise intensity of the local
noise $\xi_n(i,j)$, and $\lambda^{-1}$ is the correlation time of
the local Gaussian colored noise. Here we set $\sigma_{loc}=\sigma$
and $\lambda=0.05$. The parameter $R$ measures the noise correlation
between a pair of neurons. $\eta_n(i,j)$ is renewed at each
iteration step $n$ and for each unit $(i,j)$ in the iterated
processing according to the algorithm proposed in
Ref.~\onlinecite{Fox_1988} (see Appendix for details). In the
following discussions, we will take $\sigma$ and $R$ as controlled
parameters.

\section{Measures of neuronal dynamics}
\label{sec:3} Two measures for quantifying the observed neuronal dynamics due to the impact of noise are employed. One is the mean firing rate $\Pi$,~\cite{Rieke} and the other is the population coherence $\kappa$.~\cite{Welsh, XJ_Wang} The mean
firing rate of the neuronal network is defined as
\begin{equation}
\Pi=\left<\pi(n)\right>_T=\left<\frac{1}{N^2}\sum_{ij}\theta[u_n(i,j)-u_{th}]\right>_T,
\end{equation}
where $u_{th}=-0.2$ is the firing threshold determined by the
action potential of the Rulkov neuron. Notably, $\theta(x)$ is a
heaviside function with $\theta(x)=1$ if $x \geq 0$ and $\theta=0$
if $x<0$. The bracket $\left< \right>$ indicates the average over
the whole iteration time $T$.

To quantify the correlations of firing events in the neuronal
network, we introduce the population coherence measure
$\kappa(\tau)$.~\cite{Welsh,XJ_Wang} $\kappa(\tau)$ is defined as
the average of the local coherence $\kappa_{ij}(\tau)$ over all
the pairs of neurons, namely:
\begin{equation}
\kappa(\tau)=\frac{1}{C_N^2}\displaystyle \sum_{i,j=1; i \neq
j}^{N}\kappa_{ij}(\tau)=\frac{1}{\frac{N(N-1)}{2}}\displaystyle
\sum_{i,j=1; i \neq j}^{N}\kappa_{ij}(\tau),
\end{equation}
where
\begin{equation}
\kappa_{ij}(\tau)=\frac{\displaystyle \sum_{l=1}^m
Y_i(l)Y_j(l)}{\sqrt{\displaystyle \sum_{l=1}^mY_i(l)\displaystyle
\sum_{l=1}^mY_j(l)}}.
\end{equation}
The coherence $\kappa_{ij}(\tau)$ between any two neurons $i$ and
$j$ is measured by the cross-correlation of their spike trains at
zero time lag within a time interval $\tau$. More specifically, we
divide the full iteration time $T$ into small bins of duration
$\tau=70$ and define the two spike trains as $Y_i(l)=0$ or $1$ and
$Y_j(l)=0$ or $1$ $(l=1,2,...,m; T/m=\tau)$, whereby $Y(l)=1$ if
the onset of a spike occurred at the $l$-th time bin, otherwise
$Y(l)=0$. Sometimes, the population coherence measure
$\kappa(\tau)$ is also used to quantify the synchronization of
neuronal firings in networks.~\cite{Welsh,Gerstein} Larger
population coherence $\kappa$ corresponds to higher correlations
between neurons inside the network.

\section{Effects of correlated Gaussian noise}
\label{sec:4}

\begin{figure*}
\begin{center}
\scalebox{0.75}[0.75]{\includegraphics{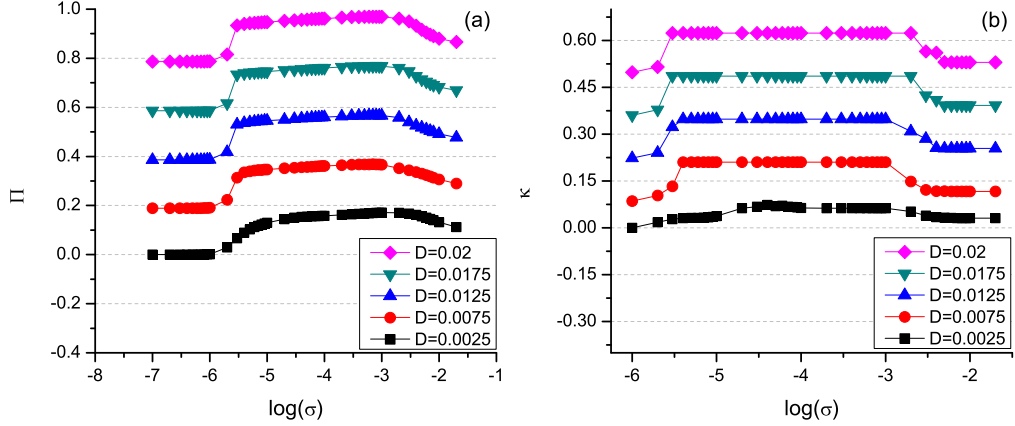}}
\end{center}
\caption{\label{fig:2} Stack lines by Y offsets of the mean firing
rate $\Pi$ (a) and the population coherence measure $\kappa$ (b)
in dependence on the noise intensity $\sigma$ of additive local
(we set $R=0.0$) Gaussian white noise for various coupling
strength $D$. Note that the $x$-axis has a logarithmic scale.}
\end{figure*}

\begin{figure*}
\begin{center}
\scalebox{0.75}[0.75]{\includegraphics{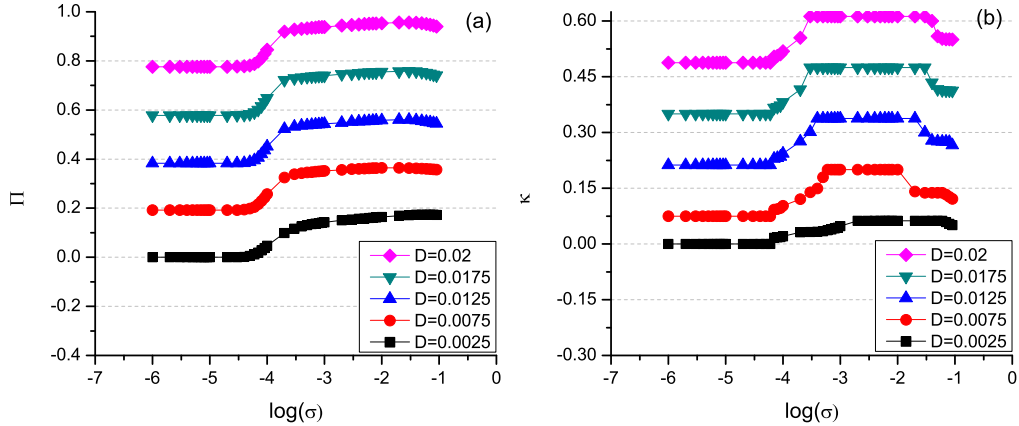}}
\end{center}
\caption{\label{fig:3} Stack lines by Y offsets of the mean firing
rate $\Pi$ (a) and the population coherence measure $\kappa$ (b)
in dependence on the noise intensity $\sigma$ of additive local
(we set $R=0.0$) Gaussian colored noise for various coupling
strength $D$. Note that the $x$-axis has a logarithmic scale.}
\end{figure*}

In order to discern clearly the distinct effects of correlated
Gaussian noise on neuronal dynamics, we consider firstly the
effects of \textit{local} Gaussian noise, i.e., $R=0.0$.
Variations of the mean firing rate $\Pi$ and the population
coherence measure $\kappa$ with respect to the noise intensity
$\sigma$ for various coupling strengthes $D$ are shown in
Figs.~\ref{fig:2} and \ref{fig:3}. Importantly, local Gaussian
noise is white in Figs.~\ref{fig:2} but colored in
Figs.~\ref{fig:3}. From these two figures, we can see that $\Pi$
and $\kappa$ can reach larger values at an intermediate noise
intensity than at smaller and larger noise intensities, for both
local Gaussian white and colored noise. We caution, however, that for large noise intensities the mean firing rate $\Pi$ may yield spurious results because of the very noisy output of individual neurons forming the neuronal network, due to which it is practically impossible to discern what is a firing event and what not. In accordance with this, the results depicted in this paper are constrained to noise intensities for which the neuronal dynamics still plays a significant role, i.e., is not completely overshadowed by noise.

The above-reported results can be interpreted as follows. Small
noise intensities are unable to evoke excitations, accordingly,
the mean firing rate $\Pi$ becomes zero and the coherence of
neurons inside the network is small (\textit{i.e.} $\kappa$ is
small). For intermediate noise intensities, typically only a few
neurons (at random) forming the lattice start firing. Due to the
diffusive coupling and the noisy support the excitations can
propagate regularly to the neighbors, which results in well
ordered circular waves (as shown in Ref.~\onlinecite{SUN_PhyA}) that
ultimately result in high firing rate $\Pi$ and large coherence
$\kappa$. While for large noise intensities, neurons inside the
network exhibit high-rate spiking behavior. When such spiking
neurons are coupled diffusively, they tend to suppress the inputs
coming from other neurons, which in turn decreases the number of
firing events in a time span and destroys the spatial order of the
dynamics, finally leading to small $\Pi$ and a decreases of
$\kappa$. Therefore, for local Gaussian noise, the mean firing
rate and the correlations of the neuronal network can be optimized
by some intermediate noise intensities, as shown in
Figs.\ref{fig:2} and \ref{fig:3}.

\begin{figure*}
\begin{center}
\scalebox{0.75}[0.75]{\includegraphics{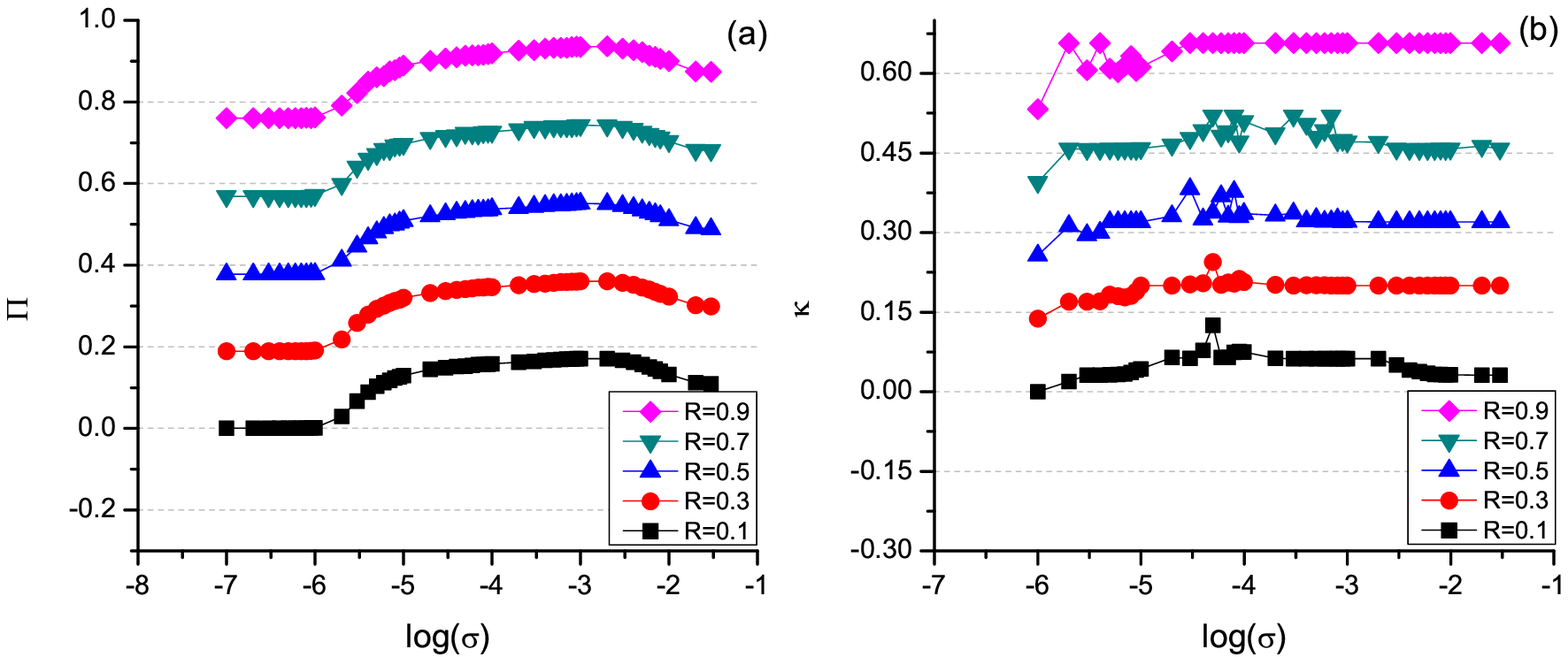}}
\end{center}
\caption{\label{fig:4} Stack lines by Y offsets of the mean firing
rate $\Pi$ (a) and the population coherence measure $\kappa$ (b)
in dependence on the noise intensity $\sigma$ of additive
correlated global noise $\eta(i,j)$ with local Gaussian white
noise for various noise correlations $R$. The coupling strength is
constant, equalling $D=0.0025$. Note that the $x$-axis has a
logarithmic scale.}
\end{figure*}

\begin{figure*}
\begin{center}
\scalebox{0.75}[0.75]{\includegraphics{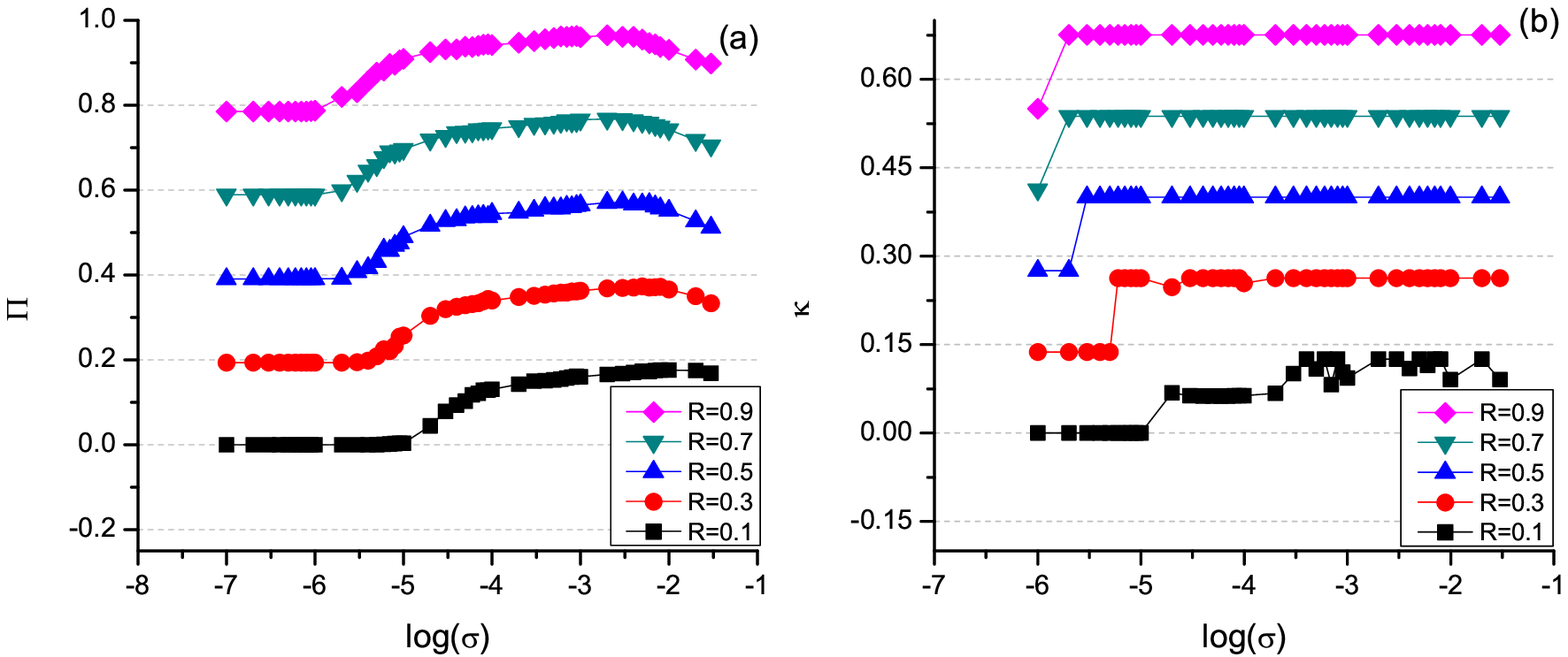}}
\end{center}
\caption{\label{fig:5} Stack lines by Y offsets of the mean firing
rate $\Pi$ (a) and the population coherence measure $\kappa$ (b)
in dependence on the noise intensity $\sigma$ of additive
correlated global noise $\eta(i,j)$ with local Gaussian colored
noise for various noise correlations $R$. The coupling strength is
constant, equalling $D=0.0025$. Note that the $x$-axis has a
logarithmic scale.}
\end{figure*}

In what follows, we examine more closely the effects of correlated
Gaussian noise via controlling the noise correlation $R$.
Variations of the mean firing rate $\Pi$ and the population
coherence measure $\kappa$ with respect to the noise intensity
$\sigma$ for various noise correlations $R$ are shown in
Figs.\ref{fig:4} and \ref{fig:5}. The local Gaussian noise is
taken as white in Figs.\ref{fig:4} and colored in
Figs.\ref{fig:5}, respectively. Compared with the results shown in
Figs.\ref{fig:2}(a) and \ref{fig:3}(a), we can see that variations
of $\Pi$ with respect to the noise intensity $\sigma$ under correlated
Gaussian noise are similar as the ones under local Gaussian noise.

For the population coherence $\kappa$, however, we can see that its
variations versus the noise intensity $\sigma$ are strongly dependent
on the noise correlation $R$ as shown in Figs.\ref{fig:4}(b) and
\ref{fig:5}(b). It is also worth pointing out that the occasional non-smoothness of the curves in the later two figures is predominantly a consequences of the somewhat erratic switching between the emergence of spatially ordered patterns and their absence on the network. The latter introduces some non-smoothness to the employed statistical quantifiers, yet we found it impossible to eliminate this by means of more intensive numerical investigations. The reader is also referred to Ref.~\onlinecite{SUN_PhyA}, where pattern formation due to correlated Gaussian noise has been studied earlier. Through comparisons with the corresponding results
presented in Figs.~\ref{fig:2}(b) and \ref{fig:3}(b), we find that
effects of correlated Gaussian noise on the population coherence
$\kappa$ look more complex than the ones of local Gaussian noise.
In the case of local Gaussian white noise, as shown in
Fig.~\ref{fig:4}(b), a weak coherent behavior (there exists a
small peak at an intermediate $\sigma$) can be observed for small $R$, e.g.,
$R=0.1, 0.3$. And for an intermediate $R$ ($R=0.5, 0.7$), the
variations of $\kappa$ become irregular with changing of the noise
intensity. With further increasing of $R$, a plateau region of
$\kappa$ emerges as the noise intensity $\sigma$ increases, as can be observed from the line corresponding to $R=0.9$, for example. For local
Gaussian colored noise, variations of $\kappa$ with respect to
$\sigma$ are irregular at $R=0.1$, and exhibit plateau regions
with increasing of $\sigma$ for large $R$, as presented in
Fig.~\ref{fig:5}(b).

\begin{figure*}
\begin{center}
\scalebox{0.63}[0.63]{\includegraphics{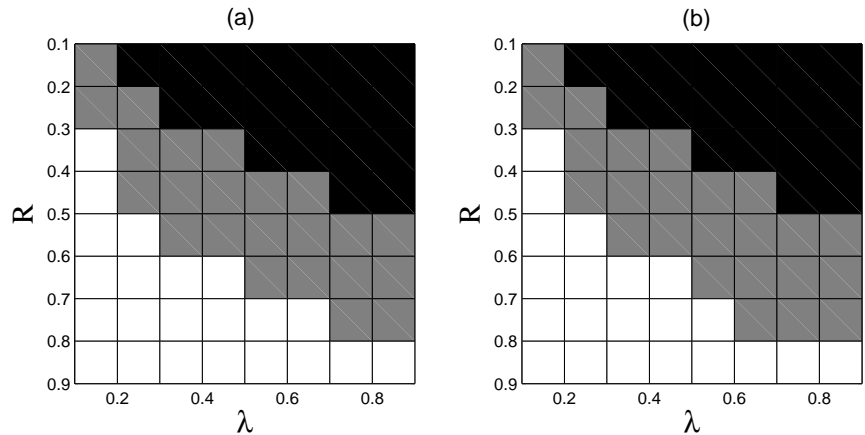}}
\end{center}
\caption{\label{fig:6} (a) Variations of $\kappa$ with respect to
$\sigma$ for different $(R,\lambda)$ pairs. If the coherence of $\kappa$ with respect to $\sigma$ is weak then the corresponding $(R,\lambda)$ pairs are colored black. If the variations of $\kappa$ with respect to $\sigma$ are irregular the color is gray, while it is white if there exists a plateau region of $\kappa$ for large $\sigma$. (b) The
dependence of $\rho$ on different $(R,\lambda)$ pairs. Color black is used for those combinations of $R$ and $\lambda$ for which $0.0 < \rho < 2.0$, gray is used if $2.0 \leq \rho < 8.0$, while white is used if $8.0 \leq \rho< \infty$. See also the main text for further details.}
\end{figure*}

Notably, the underlining mechanism regarding the effects of correlated Gaussian noise on the mean firing rate $\Pi$ is similar as discussed above in the case of local Gaussian
noise. However, in order to clarify the effects of noise correlation $R$ on the complicated variations of $\kappa$ versus $\sigma$, we introduce the quantity
\begin{equation}
\rho=\frac{2R\sigma}{(1-R)\langle\xi(i,j)\rangle^2},
\end{equation}
which is the noise strength ratio between global and local noise.
$\rho=R/(1-R)$ when local noise is taken as Gaussian white noise,
and $\rho=2R/(1-R)\lambda$ when it is taken as Gaussian colored
noise. Thus, values of $\rho$ can be controlled by the noise
correlation $R$ and $\lambda$. We calculate the variations of
$\kappa$ with respect to $\sigma$ for different noise intensity
ratios $\rho$, or equivalently, different pairs of $(R,\lambda)$. The obtained
results are presented in Fig.~\ref{fig:6}(a), where pairs
of $R$ and $\lambda$ are colored black if the variation of
$\kappa$ with respect to $\sigma$ shows a weak coherence. At this point it is instructive to examine the corresponding curves for $R=0.1$ and $0.3$ that are depicted in Fig.~\ref{fig:4}(b). Conversely, pairs of $R$ and $\lambda$ are colored gray if
variations of $\kappa$ with respect to $\sigma$ are irregular, as can be observed
observed from the corresponding curves for $R=0.5$ and $0.7$ in
Fig.~\ref{fig:4}(b) and from the curve depicted for $R=0.1$ in Fig.~\ref{fig:5}(b). Finally, $(R,\lambda)$ pairs are colored white if there exists
a plateau region of $\kappa$ for large $\sigma$, as can be observed in Fig.~\ref{fig:4}(b) for $R=0.9$ and in Fig.~\ref{fig:5}(b) for $R=0.3$, $0.5$, $0.7$ and $0.9$. In order to now appreciate the quantity $\rho$ introduced in Eq.~(8) as an important driving force behind the variations $\kappa$ with respect to $\sigma$, we show in Fig.~\ref{fig:6}(b) how $\rho$ varies for different pairs of $R$ and $\lambda$ in a systematic manner. In particular, from thus far presented results it can be concluded that there exist three intervals of $\rho$, \textit{i.e.} $(0,a), [a,b), [b, \infty)$,
within which the variations of $\kappa$ with respect to $\sigma$
are different from one another. In Fig.~\ref{fig:6}(b) black is used for those combinations of $R$ and $\lambda$ for which $0.0 < \rho < 2.0$, gray is used if $2.0 \leq \rho < 8.0$, while white is used if $8.0 \leq \rho< \infty$. Compared to results presented in Fig.~\ref{fig:6}(a), we can observe at a glance that by setting $a$ and
$b$ to equal $2.0$ and $8.0$, respectively, the color patterns match nearly perfectly, from which we conclude that $\rho$ indeed has a decisive impact on variations of $\kappa$ with respect to $\sigma$.

From the analysis of the effects of correlated Gaussian noise, we
now thus know that noise correlations have no notable effects on the
variations of $\Pi$ versus $\sigma$, while conversely, the noise correlation
$R$ plays a crucial role in how $\kappa$ varies with
respect to $\sigma$, in particular by means of controlling the noise strength ratio
$\rho$.

\section{Summary}
\label{sec:5}

In this paper, we have studied effects of correlated Gaussian
noise on neuronal firings, measured by the mean firing rate and
the population coherence, of a two-dimensional network, which is
locally modeled by the Rulkov map. Based on our numerical
simulations, we have found that the mean firing rate of the
network $\Pi$ can be enhanced at some intermediate noise
intensities by correlated Gaussian noise for any noise correlation
$R$. This phenomenon is similar as the coherence resonance, even
though there is not a well-defined optimal noise intensity. While
for population coherence measure $\kappa$, we find that its
variations with respect to noise intensity are very complex.
Furthermore, we reveal that it strongly depends on the ratio
$\rho$. Moreover, through the measure for population coherence
$\kappa$, we have been able to gain a deeper understanding of
the interactions between global and local noise. In particular, we
have shown that an appropriately tuned global noise can be an
effective promotor of correlations of firing events in the
neuronal network.

As we have already stated in the Introduction, it is still debatable whether a cortical
neuron is driven mainly by the mean firing rate of presynaptic
neurons or by correlations between pairs of neurons inside the
neuronal network. Moreover, neurons inside neuronal networks are
not only affected by local random fluctuations, but also
stimulated by some common random inputs. Thus, the results
obtained for discussing effects of correlated Gaussian noise on
the mean firing rate and correlations of the neuronal network may
give some important implications on investigating transmission of
neuronal information in neuronal networks.

\begin{acknowledgments}
This work was supported by the National Natural Science Foundation
of China (grants 10872014 and 10972018). XS is acknowledges funding from the Chinese Postdoctoral Science Foundation (grant 20090460337). MP thanks the Slovenian Research Agency (grant Z1-2032). JK acknowledges support form the German Science Foundation
(program SFB 555).
\end{acknowledgments}

\appendix*
\section{Noise generation}
Gaussian white noise $e_n$ can be generated effectively for map-based neuronal networks as follows. Let
\begin{equation}
\begin{array}{ll}
a=\rm{random~~number}, \\
b=\rm{random~~number},\\
e_n=[-4\sigma \ln(a)]^{1/2}\cos(2\pi b),
\end{array}
\end{equation}
where $a$ and $b$ are uniformly distributed on the unit interval. When $\xi_n(i,j)$ is Gaussian white noise, $e_n$ will be renewed according to Eq.~(A.1) at each iteration step $n$ and for each unit $(i,j)$. When $\xi_n(i,j)$ is a Gaussian colored noise, however, it can be generated by means of
\begin{equation}
\begin{array}{ll}
a=\rm{random~~number}, \\
b=\rm{random~~number},\\
g_w=[-4\sigma\Delta t \ln(a)]^{1/2}\cos(2\pi b),\\
\xi_{n+1}=\xi_{n}- \lambda \xi_{n}+\lambda g_w,
\end{array}
\end{equation}
with initial conditions
\begin{equation}
\begin{array}{ll}
l=\rm{random~~number}, \\
m=\rm{random~~number},\\
g_c=[-4\sigma \lambda \ln(l)]^{1/2}\cos(2\pi m),\\
\end{array}
\end{equation}
where $a,b,l$ and $m$ are uniformly distributed on the unit interval.
Subsequently, $\eta_n(i,j)=\sqrt{R}e_n+\sqrt{1-R}\xi_n(i,j)$ is renewed at each
iteration step $n$ and for each unit $(i,j)$ by repeating Eqs.~(A.2) and (A.3).

\end{document}